\documentclass[aps,showpacs,tightenlines]{revtex4}
\usepackage[dvips]{graphicx}
\usepackage{subfigure}
\usepackage[english]{babel}
\usepackage[T1]{fontenc}
\usepackage{mathrsfs}
\usepackage[tbtags]{amsmath}
\usepackage{amssymb}
\usepackage{amsopn}
\usepackage{latexsym}
\usepackage[mathcal]{eucal}
\usepackage[flushleft,font={small},labelfont=bf]{caption}
\usepackage{color}

\newcommand{\beq}{\begin{equation}}
\newcommand{\eeq}{\end{equation}}
\newcommand{\crea}[1]{c^{\dagger}_{\sigma}(\vec{#1})}
\newcommand{\ann}[1]{c_{\sigma}(\vec{#1})}

\newcommand{\sumr}{\sum_{\vec{r}}}

\newcommand{\dpsi}{D[\bar{\Psi}\Psi]}
\newcommand{\dphi}{D[\vec{\phi}\>]}
\newcommand{\dphip}{D[\vec{\phi}\>'\>]}
\newcommand{\coupling}{\frac{U}{t}}

\DeclareMathOperator{\hc}{h.c.}

\DeclareMathOperator{\Det}{Det}
\DeclareMathOperator{\tr}{Tr}

\begin{document}
\title[GEP and antiferromagnetism in the Hubbard model]{Gaussian effective potential and antiferromagnetism \\
in the Hubbard model}
\author{L. Marotta and F. Siringo}
\affiliation{Dipartimento di Fisica e Astronomia, 
Universit\`a di Catania,\\
INFN Sezione di Catania and CNISM Sezione di Catania,\\
Via S.Sofia 64, I-95123 Catania, Italy}
\begin{abstract}
The Gaussian Effective Potential (GEP) is shown to be a useful variational
tool for the study of the magnetic properties of strongly correlated electronic
systems. The GEP is derived for a single band Hubbard model on a two-dimensional
bi-partite square lattice in the strong coupling regime. At half-filling the  
antiferromagnetic order parameter emerges as the minimum of the effective potential
with an accuracy which improves over RPA calculations and is very close to that
achieved by Monte Carlo simulations. Extensions to other magnetic systems are discussed.
\end{abstract}
\maketitle
\section {Introduction}
The Gaussian Effective Potential (GEP) is a well established variational tool
~\cite{schiff,rosen,barnes,kuti,chang,weinstein,huang,bardeen,peskin,stevenson}
which has been mainly used and developed for describing 
the breaking of symmetry in the standard model of
electroweak interactions~\cite{stevenson,ibanez,siringo_sigma,siringo_06,
siringo_su2,siringo_bubble}.
Quite recently, the GEP has also been shown to provide a very useful non-perturbative
method for the study of condensed matter phenomena like 
superconductivity~\cite{camarda,SC_gep,marotta} 
and we expect that many other interesting
phenomenological aspects could be described by the same tool.
Quite generally, the method allows to study the broken-symmetry ground state of
a bosonic theory describing quantum and thermal fluctuations 
for a large class of physical systems. Actually, a Bose field emerges in the
theoretical description of many physical systems
like superconductors, superfluids, magnetic materials, disordered fermions and
many others. Thus we believe that the whole capabilities of the GEP 
have not been explored yet.

In this paper we show that the GEP can be used for describing the strong-coupling
limit of magnetic materials, where a spontaneous breaking of symmetry occurs at
low temperature, and the local order parameter (i.e. the local magnetic moment)
is a Bose field with excited states known as spin waves.
In the strong-coupling limit, the usual perturbative methods are not reliable and
the more important features are usually recovered by other techniques
like RPA\cite{RPA}, fluctuation exchange (FLEX) approximation \cite{Flex}, functional 
renormalization group (FRG) \cite{RG} and numerical Monte Carlo 
calculations \cite{MC}. When there are no exact
results available, Monte Carlo outputs are often regarded like experimental data even if
such numerical calculations are plagued by several shortcomings like border effects,
small size of the samples and a finite temperature. We show that the GEP provides
an analytical variational tool for describing infinite systems at zero temperature 
with an accuracy which improves over RPA and is close to that achieved by Monte Carlo.

In order to compare the results we study the well 
known half-filling Hubbard model~\cite{hubbard}
in two spatial dimensions on a square lattice. The model has a broken-symmetry
magnetic ground state at any coupling, and in the strong-coupling limit
the local magnetic moment is known to saturate at a smaller value than predicted by
the simple mean-field theory. According to Monte Carlo data~\cite{MC} and RPA
calculations~\cite{RPA} the local moment is reduced by almost one half as a consequence
of quantum fluctuations, in agreement with the theoretical limit which can
be extracted from the Heisenberg model~\cite{heisenberg_sw,heisenberg_mc}.
The GEP predicts the correct strong-coupling limit of the Hubbard model, 
and we believe that such a variational method may provide a useful analytical
interpolation for any coupling. Unfortunately we were not able to get the 
GEP exactly from the Hubbard model, and we had to approximate the exact spin wave
Lagrangian by a power expansion. Neglecting higher order powers can be shown
to be reasonable in the strong coupling limit, but there is no control of the
approximation in the small-coupling regime and eventually at the quantum transition
where fluctuations become very large. Thus we could not check the validity of
the GEP in the small-coupling regime where, on the other hand, 
mean-field and perturbative methods
are known to work well. That shortcoming is a limit of the present calculation 
while we expect that the exact GEP should give the correct prediction for
any coupling.
However the calculation has its merits and 
shows the potentiality of the GEP 
in the non-perturbative strong-coupling regime where alternative 
analytical methods are particularly welcome. Once the method has been
tested on the well known Hubbard model, and its limits are understood,
we believe that the GEP may represent a valid alternative tool for the
study of other more complex magnetic systems.

The paper is organized as follows: in the next section 
the GEP approximation for the Hubbard model is discussed; then the 
effective potential is evaluated at half-filling in section \ref{sectionIII} where 
the phenomenological predictions of the method are 
compared with the available Monte Carlo data.
 
\section {The Hubbard model: definitions and GEP approximation}
\label{sectionII}

Let us define the single-band Hubbard model on a two-dimensional bipartite 
square lattice according to the Hamiltonian
\beq
H=-t\sum_{\substack{\langle \vec{r},\vec{r}'\rangle \\ 
\sigma=\uparrow,\downarrow}}
\left(\crea{r}\ann{r'}+\hc\right)+
U'\sum_{\vec{r}}n_{\uparrow}(\vec{r})n_{\downarrow}(\vec{r}).
\label{Hubb_H}
\eeq
where $\crea{r},\ann{r}$ are the usual fermionic creation and annihilation operators,
$\langle,\rangle$ restricts the sum to nearest neighbour sites, and $n_{\sigma}(\vec{r})=\crea{r}\ann{r}$ 
is the particle number operator at site $\vec{r}$ for spin $\sigma$ 
(Below we set $U$=$U'/3$ in order to agree with a widely used notation\cite{MC,RPA}).

According to the standard path-integral approach\cite{fradkin},  
in the basis of coherent states for 
fermions, the fluctuation amplitude $Z$ reads:
\beq
Z=\int \dpsi e^{i\int dt\>L}
\eeq
where $\Psi,\bar{\Psi}$ are the Grassmann variables associated with the 
fermionic annihilation and creation operators and $L$ is the Lagrangian density.
An auxiliary bose field  $\vec{\phi}$ is introduced at any space-time point
by the Hubbard-Stratonovich transformation, and the  
Grassmann fields can be integrated out exactly yielding 
\beq
Z=\int\dphi e^{iS_{eff}(\vec{\phi})}, 
\eeq
with the effective action
\beq
S_{eff}(\vec{\phi})=-\frac{1}{2}\int dt
\sumr\vec{\phi}^2(\vec{r},t)
-i\ln\Det\left(i\partial_t+\mu-M(\vec{\phi})\right)
\label{Seff}
\eeq
written in terms of the matrix $M(\vec{\phi})$ 
\beq
\langle\vec{r},t,\alpha|M(\vec{\phi})|\vec{r}{\>'},t',\beta\rangle =\delta_{t,t'}
\left[-2t\delta_{\alpha,\beta}\sum_{\langle\vec{r}''\rangle}
\delta_{\vec{r}'',\vec{r}'}+\sqrt{U}\vec{\phi}
(\vec{r},t)\cdot\vec{\tau}_{\alpha,\beta}\delta_{\vec{r},\vec{r}'}\right].
\label{M}
\eeq
where $\vec{\tau}_{\alpha\beta}$ are the Pauli matrices.
The bosonic effective action $S_{eff}$ is exact, 
and describes the spin-wave
excited states of the Hubbard model at zero temperature\cite{fradkin2}.

We would like to build the GEP for the effective action (\ref{Seff}),
and as a first step we take a shift of the bosonic field and write
$\vec{\phi}\equiv\vec{\phi}(\vec{r},t)$ as the sum of a 
background constant (non homogeneous) 
field $\vec{\phi}_0$ plus a fluctuating field $\vec{\phi\>'}$
\beq
\vec{\phi}(\vec r,t)=\vec{\phi}_0(\vec r)+\vec{\phi\>'}(\vec r,t).
\eeq

The effective action becomes
\beq
S_{eff}[\vec{\phi}_0+\vec{\phi\>'}]=
-\int dt\sumr \frac{1}{2}\left(\vec{\phi}_0+\vec{\phi\>'}\right)^2
-i\ln\Det\left(i\partial_t+\mu-M(\vec{\phi}_0)-\delta M(\vec{\phi\>'})\right)
\label{Seff2}
\eeq
where the matrix $\delta M$ 
is defined according to
\beq
\langle\vec{r},t,\alpha|\delta M(\vec{\phi\>'})|\vec{r}{\>'},t',\beta\rangle
=\delta_{t,t'}\sqrt{U}
\vec{\phi\>'}\cdot\vec{\tau}_{\alpha,\beta}\delta_{\vec{r},\vec{r}'}.
\label{deltaM}
\eeq

Then we can split the effective action as $S_{eff}=S_0+S_{int}$ and write
\begin{align}
S_0&=-\frac{1}{2}\int dt\sumr{\vec{\phi}_0}\>^2+i\tr\ln\left(iG_0(\vec{\phi}_0)\right)\\
S_{int}&=-\int dt\sumr{\vec{\phi}_0}{\vec{\phi\>'}}
-\frac{1}{2}\int dt\sumr{\vec{\phi\>'}}\>^2+ \notag\\
\quad &\quad+\sum_n\frac{i^{n+1}}{n}
\tr \left(iG_0(\vec{\phi}_0)\delta M(\vec{\phi\>'})\right)^n
\label{expansion}
\end{align}
where $iG_0(\vec{\phi}_0)$ is the mean-field fermion Green function
in a background 
magnetic field $\vec{\phi}_0$:
\beq
iG_0(\vec{\phi}_0)={\left(i\partial_t+\mu-M(\vec{\phi}_0)\right)^{-1}}.
\eeq
The expansion is exact provided that all the infinite terms are included.

It is important to note 
that symbols here have a different meaning compared to the standard expansion reported
in textbooks\cite{fradkin}.
Actually, in our calculation the 
background field $\vec{\phi_0}$ is not fixed at the saddle-point of the 
effective action in \eqref{Seff}, but it represents a variational parameter 
to be evaluated by a minimization of the Gaussian effective potential
which includes quantum fluctuations.
Similarly, the shifted field $\vec{\phi'}$ does not describe just the 
fluctuations around the mean--field value of $\vec{\phi_0}$: it embodies 
the behaviour of quantum fluctuations for every different value 
of the background field. Thus in this variational calculation 
a different quantitative result is obtained by use of the basic standard
formalism.

Let us introduce the trial Gaussian action term $S_{GEP}$
\beq
S_{GEP}=\frac{1}{2}\sum_{\vec r,\vec r^\prime}
\int dt\phi^\prime_{a}(\vec{r},t){(g^{-1})^{ab}}\phi^\prime_{b}(\vec{r}\>',t')
\eeq
with the matrix elements $g_{ab}(\vec{r},t,\vec{r}\>',t')$ playing the role
of variational parameters, and let us write the effective action as 
\beq
S_{eff}=S_{GEP}+\left(S_0+S_{int}-S_{GEP}\right).
\eeq
According the amplitude $Z$ reads
\beq
Z[\vec{\phi_0}]=\int \dphip e^{iS_{GEP}}e^{i(S_0+S_{int}-S_{GEP})}
\eeq
and we can write
\beq
\ln Z[\vec{\phi_0}\>]=
\ln Z_0[\vec{\phi_0}\>]+
\ln\langle e^{i(S_0+S_{int}-S_{GEP})}\rangle_{GEP}
\eeq
where
\beq
Z_0[\vec{\phi_0}\>]=\int \dphip e^{iS_{GEP}}
\label{Z0}
\eeq
and with the Gaussian averages defined as
\beq
\langle(\ldots)\rangle_{GEP}=
\frac{\int \dphip (\ldots)e^{iS_{GEP}}}{\int \dphip e^{iS_{GEP}}}.
\label{average}
\eeq
As usual~\cite{stevenson,camarda,marotta}, the GEP is defined according to
\beq
V_{GEP}[\vec{\phi_0}\>]=
\frac{1}{\cal V}\left(i\ln Z_0[\vec{\phi_0}\>]
-\langle S_0+S_{int}-S_{GEP}\rangle_{GEP}\right). 
\label{eff_pot}
\eeq
where ${\cal V}$ is a total space-time volume.
The effective potential $V_{GEP}$ is the energy density of the system, and it
obviously depends on the choice of the matrix elements $g_{ab}$ that will be
regarded as variational parameters.
The evaluation of the GEP then only requires the calculation of the
gaussian averages in (\ref{eff_pot}) according to Wick's theorem with
\beq
\langle\phi'_{a}(\vec{r},t)\phi'_{b}
(\vec{r}\>',t)\rangle_{GEP}=ig_{ab}(\vec{r},t,\vec{r}\>',t').
\label{gep_green}
\eeq
Unfortunately the expansion in (\ref{expansion}) can only be evaluated up
to some finite order, which means that some extra approximation is added besides
the genuine variational GEP of (\ref{eff_pot}). 
Neglecting higher order powers is a very reasonable approximation
in the strong coupling limit, but is expected to fail in the small coupling regime.
In fact the approximation may be regarded as an expansion in powers of the 
fluctuating field $\vec \phi\>'$ around the generic point $\vec{\phi_0}$:
at any finite order, (\ref{expansion}) fits the exact effective action 
around $\vec{\phi_0}$ in a neighbourhood 
that will get larger as more terms are added 
to the expansion. According to (\ref{eff_pot}), the GEP
is evaluated by a sampling of $S_{int}[\vec{\phi\>'}]$ 
through the Gaussian functional by (\ref{average}), then
the accuracy of the GEP depends 
on the width of the Gaussian functional $\exp(iS_{GEP})$: 
a narrow trial functional will average 
the approximate effective action around $\vec{\phi_0}$ on a small range where 
the expansion is a good fit of the exact effective action;
a broader functional will average the effective action on 
a larger range where the expansion departs from the exact action, 
resulting in a totally uncontrolled approximation and wrong predictions.
But the width of the wave functional is determined by the relative weight
of fluctuations which, in turn, are related to the Coulomb coupling $U$; actually, 
it is well known that large values of the coupling reduce fluctuations
of the local moment (\emph{i.e.} the Hubbard-Stratonovich bose field). 
In this sense, we expect that in the strong coupling limit even a second order 
expansion in (\ref{expansion}) should be enough while in the weak coupling 
regime and at the quantum transition,
where fluctuations become very large, the method is doomed to fail. 
However, in principle, the method could be improved 
order by order, even if the real calculation 
of the gap equation is expected to become very difficult.

With such limitations understood, we can easily evaluate the GEP by use of the
simple second order approximation, and content ourselves with the strong coupling
regime of the model.
Neglecting costant terms, we can write
\beq
\ln Z_0[\vec{\phi_0}\>]
=\frac{1}{2}\tr\ln g
\eeq
and the averages read
\beq
\langle S_0\rangle_{GEP}=-\frac{1}{2}\int dt\sumr{\vec{\phi}_0}\>^2(\vec{r})
+i\tr\ln\left(iG_0(\vec{\phi}_0)\right)
\eeq
\beq
\langle S_{int}\rangle_{GEP}\approx
-\frac{1}{2}\int dt \sumr\langle{\vec{\phi}\>'}^2(\vec{r},t)\rangle_{GEP}
-\frac{i}{2}\tr \langle\left(iG_0(\vec{\phi}_0)\delta M(\vec{\phi}\>')
\right)^2\rangle_{GEP}
\eeq

where we have summed up to second order and have omitted odd terms whose 
average is zero.
We notice that the variational parameters $g_{ab}$ are implicit functionals
of the local background field $\vec \phi_0(\vec r)$, as they are going to be determined
for any local configuration of the field by requiring that the effective potential
is stationary.
Using \eqref{deltaM} and \eqref{gep_green} we obtain
\beq
\begin{split}
\tr \langle\left(iG_0(\vec{\phi}_0)\delta M(\vec{\phi}\>')
\right)^2\rangle_{GEP}=iU\int dt&\int dt'\sum_{\vec r,\vec r'}
\big({G_0}_{\alpha\alpha'}
(\vec{r},t;\vec{r}\>',t',\vec{\phi}_0)\tau^a_{\alpha'\beta}\times\\
&\quad\times{G_0}_{\beta\beta'}(\vec{r}\>',t';\vec{r},t,\vec{\phi}_0)
\tau^b_{\beta'\alpha}g_{ab}(\vec{r}\>',t';\vec{r},t)\big).
\end{split}
\eeq
Finally we can Fourier transform and according to (\ref{eff_pot})
we are left with the effective potential
\beq
\begin{split}
V_{GEP}[\vec{\phi}_0]&=\frac{i}{2}\sum_a\int_{\vec{q},\Omega}\ln g_{aa}
(\vec{q},\Omega)-i\sum_\alpha\int_{\vec{q},\Omega}\ln \left(i{G_0}_{\alpha\alpha}
(\vec{q},\Omega,\vec{\phi}_0)\right)+
\frac{1}{2{\cal V}}\int dt \sumr\vec{\phi}_0\>^2(\vec r) + \\
&\quad+\frac{i}{2}\sum_a\int_{\vec{q},\Omega} g_{aa}(\vec{q},\Omega)
-\frac{U}{2}\int_{\vec{q},\Omega}g_{ab}(\vec{q},\Omega)K^{ab}(\vec{q},\Omega)
\end{split}
\eeq
where $\int_{\vec{q}, \Omega}\equiv\int\frac{d^2\vec{q}}{(2\pi)^2}\frac{d \Omega}{2\pi}$ and
\beq
K^{ab}(\vec{q},\Omega)={\tr}
\int_{\vec{k},\omega}G_0(\vec{k},\omega,\vec{\phi}_0)\tau^a
G_0(\vec{k}-\vec{q},\omega-\Omega, \vec{\phi}_0)\tau^b.
\label{K}
\eeq

For any local configuration of the background field $\vec \phi_0(\vec r)$
the matrix $g$ follows from the stationary condition
$\delta V_{GEP}/{\delta g_{ab}}=0$ (gap equation) which reads
\beq
(g^{-1})^{ab}=-\delta^{ab}-iUK^{ab}.
\label{gap}
\eeq
This gap equation defines the matrix elements $g_{ab}$ as implicit functionals
of the field $\vec \phi_0(\vec r)$,
and by insertion in (\ref{eff_pot}) the GEP becomes
\beq
V_{GEP}[\vec{\phi}_0]=\frac{i}{2}\sum_a\int_{\vec{q},\Omega}\ln g^{aa}
(\vec{q},\Omega)-i\sum_\alpha\int_{\vec{q},\Omega}\ln \left(i{G_0}_{\alpha\alpha}
(\vec{q},\Omega,\vec{\phi}_0)\right)+
\frac{1}{2{\cal V}}\int dt \sumr\vec{\phi}_0\>^2(\vec r).
\label{GEP}
\eeq
where we have omitted all constant terms. 
The three terms in (\ref{GEP}) are easily recognized as the quantum 
spin fluctuation energy, the electron energy and the classical spin energy
respectively. 

A closer inspection of \eqref{gap} allows us to obtain a deeper
understanding about the nature of our approximation and its 
relationship with other approaches to the Hubbard model. From a formal point 
of view, \eqref{gap} tells us that the spin correlation matrix $\hat{g}$ 
can be seen as the sum of a geometric series whose common ratio is 
$iu\hat{K}$. This leads to a diagrammatic expansion of $\hat{g}$ in terms 
of bubble and ladder graphs typically involved in the RPA calculation of 
quantum fluctuations. 
So, the GEP does not introduce the resummation of any
different class of diagrams as, \emph{e.g.}, the FLEX approximation ~\cite{Flex}.
Anyway, RPA and GEP are not equivalent. In fact, in the 
random phase approximation the evaluation of quantum corrections to the sublattice 
magnetization makes use of the one-particle Green's function obtained from the 
mean-field value of the background field $\vec{\phi_0}$ ~\cite{RPA}. In the GEP, instead, as 
already pointed out above, the background field dependence of the fluctuation matrix 
relies on a variational minimization, so avoiding to fix $\vec{\phi_0}$ at its mean-field value.
Furthermore, the reliability of our approach in the 
strong and very strong coupling limit makes it a complementary tool to 
RG methods like the Interaction Flow scheme \cite{RG}, 
which become inaccurate when the interaction is equal or greater
than the bandwidth, but works very well for small values of $U/t$. 

The ground state magnetization is determined by the stationary condition of
the effective potential $\delta V_{GEP}/{\delta \vec{\phi_0}}=0$.
The mean-field result is recovered by neglecting the quantum fluctuations:
in fact if we neglect the first term on the right hand side of (\ref{GEP})
the stationary condition reads
\beq
\vec \phi_0 (\vec r)=-\sqrt{U} \tr\left[\vec \tau G_0(\vec\phi_0)\right]
\label{MF_phi}
\eeq
which is the standard mean-field self-consistency equation.
The functional dependence of the matrix elements $g_{ab}$ on $\vec \phi_0$  ensures
that the minimum of the total effective potential occurs at a shifted value with
respect to the mean-field magnetization. 

 Furthermore, \eqref{MF_phi} establishes a relation between the 
fluctuation field and the sublattice magnetization. In fact, since 
the local mean-field magnetic moment is equal to
\beq
\langle\vec{S}(\vec{r},t)\rangle=\tr\left[\frac{\vec\tau}{2} G_0(\vec\phi_0)\right],
\eeq
where the spin operator is
\beq
\vec{S}(\vec{r})=\crea{r}\frac{\vec{\tau}_{\sigma\sigma'}}{2}\ann{r},
\eeq
we can easily write
\beq
\langle\vec{S}(\vec{r},t)\rangle=-\frac{1}{2}\sqrt{\frac{1}{U}}\vec \phi_0 (\vec r).
\eeq
This relation is more general than mean-field approximation 
and can be also recovered from the
definition of the field $\vec \phi$ (\cite{fradkin}).
Keeping in mind this result, we will consider the effective potential $V_{GEP}$ as a 
function of the magnetization $m$=2$\vert\langle\vec{S}(\vec{r},t)\rangle\vert$.

\section{Results at half filling}
\label{sectionIII}
The half-filled Hubbard model is known to be an antiferromagnet
for any strength of the coupling $U$. We assume a staggered  
magnetization on the bipartite square lattice
\beq
\vec \phi_0 (\vec r)=\pm \varphi \vec n
\eeq
where $\vec n$ is any constant unimodular versor, 
$\vert \vec n\vert=1$, and the sign is positive on a sub-lattice
and negative on the other one. The effective potential becomes
a function of the scalar $\varphi$ which gives the strength of the
local magnetization. The versor $\vec n$ breaks the rotational
symmetry of the model, and the opposite signs of $\vec \phi_0$ on
the sub-lattices break the translational symmetry of the square
lattice. The unit cell doubles its size as it is replaced by the
unit cell of the sub-lattice. The first Brillouin zone reduces
to one half and its boundary is at the Fermi surface of the
unperturbed electron gas at half filling (perfect nesting).
Let us define the two-component Green function $G^{\pm}(\varphi)$
\beq
i {G^\pm }_{\alpha\beta} (\vec k, \omega;\varphi)=
\frac{\left[\omega\pm\epsilon(\vec k)\right]\delta_{\alpha\beta}
+\sqrt{U}\varphi\>\vec n\cdot\vec \tau_{\alpha\beta}}
{\omega^2-E\>^2(\vec k)+i\eta}
\label{Green_f}
\eeq
where the free-electron band energy $\epsilon(\vec k)$
is 
\beq
\epsilon (\vec k)=-2t\sum_a\cos(k_a)
\eeq
and 
the mean field band reads
\beq
E(\vec k)=\sqrt{\epsilon^2(\vec k)+U\varphi^2}.
\eeq
\begin{figure}[htb]
\begin{center}
\subfigure[]{\label{fig:subfig:a}\includegraphics[height=7cm,width=7cm]{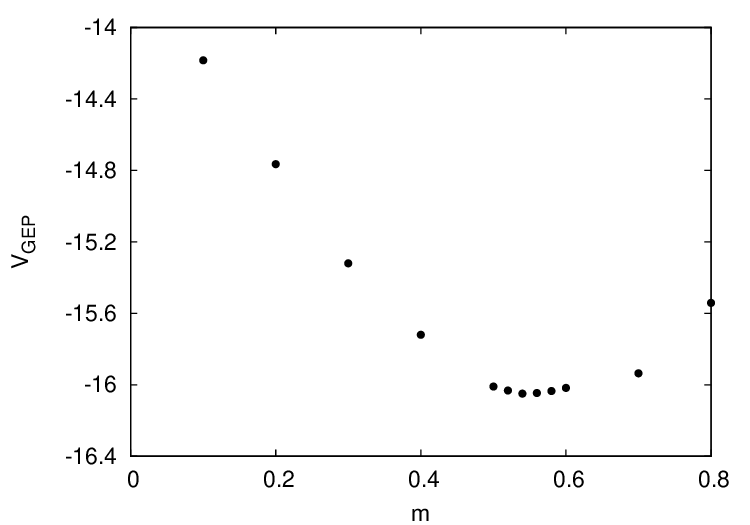}}
\hspace{1cm}
\subfigure[]{\label{fig:subfig:b}\includegraphics[height=7cm,width=7cm]{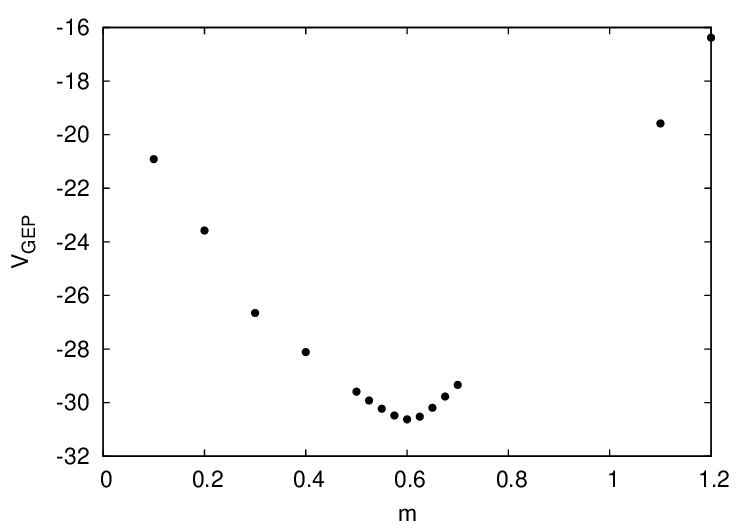}}
\caption{Effective potentials in units of the bandwidth $t$ for $\coupling$=10 (a) 
and $\coupling$=30 (b). The minima of the effective potentials, i.e. the ground state 
magnetization, occur for $m$=0.55 in (a) and $m$=0.6 in (b) (see further details in the text).
}
\label{fig:1}
\end{center}
\end{figure}

At half filling ($\mu=0$) the electron
Green function $G_0(\varphi)$ reads\cite{fradkin}
\beq
G_0(\vec k,\omega; \varphi)=G^+(\vec k,\omega; \varphi)
\eeq
for any $\vec k$ belonging to the reduced first zone
(below the unperturbed Fermi energy), while outside the
reduced first zone (above the unperturbed Fermi energy)
the Green function is given by the second component
\beq
G_0(\vec k,\omega; \varphi)=G^-(\vec k,\omega; \varphi).
\eeq
Insertion into (\ref{K}) and (\ref{gap}) yields the
spin wave correlation matrix $g$. 

The $\omega$ integration in \eqref{K} can be carried out 
analytically. In this way we are 
left with a two-dimensional k-integration for the real 
part of $g$ and a unidimensional one for its imaginary part, both 
to be performed by numerical methods.

The real part has been obtained integrating by Simpson's rule
on a $70\times70$ 
grid in the full first Brillouin zone, while for the imaginary part we 
used a $3000$ point sampling of the interval $[-\pi,\pi]$.

The effective potential $V_{GEP}$ can now be evaluated 
through \eqref{GEP}. Exploiting the parity 
properties of $g(\vec{q},\Omega)$ the first integral 
in \eqref{GEP} can be restricted to positive values 
of $\Omega$ and to the upper right quarter of the 
Brillouin zone. The employed grid contains 
$200\times20\times20$ points in the 
$(\Omega\times\vec{q})$ space.

Figure \ref{fig:1} presents two plots of the effective potential 
in units of the bandwidth $t$ for $\coupling$=10 (figure \ref{fig:subfig:a})
and $\coupling$=30 (figure \ref{fig:subfig:b}). As expected, the effective potential 
shows a minimum for a non vanishing value of $m$, which corresponds to 
the magnetization of the broken symmetry ground state; 
a more accurate estimate of $m$ is then obtained by a least square fit of 
the region of the minimum with a parabolic curve.

\begin{figure}[htb]
\begin{center}
\includegraphics[height=8cm,width=8cm]{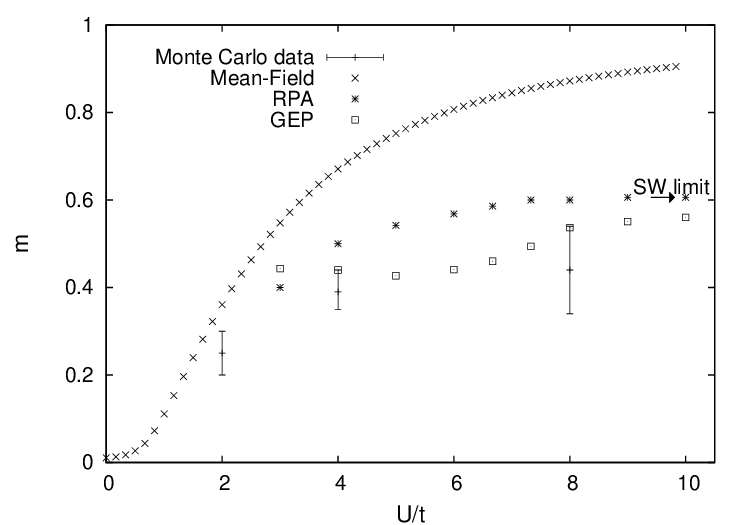}
\caption{Sublattice magnetization $m$ vs $\coupling$ within 
the GEP  (blank squares), RPA (stars), Monte Carlo (errorbars) 
and mean-field (crosses) approximations. The arrow indicates the spin-wave 
strong coupling limit m=0.606.
}
\label{fig:2}
\end{center}
\end{figure}

Another important feature of our approximation can be observed 
in figure \ref{fig:1}: the potential broadens as $\coupling$ decreases. 
This behaviour implies that the smaller is the coupling the larger are 
the fluctuations of the field, so we expect that the simple second order 
expansion of \eqref{expansion} should fail in the small coupling regime.
\begin{figure}[htb]
\begin{center}
\includegraphics[height=8cm,width=8cm]{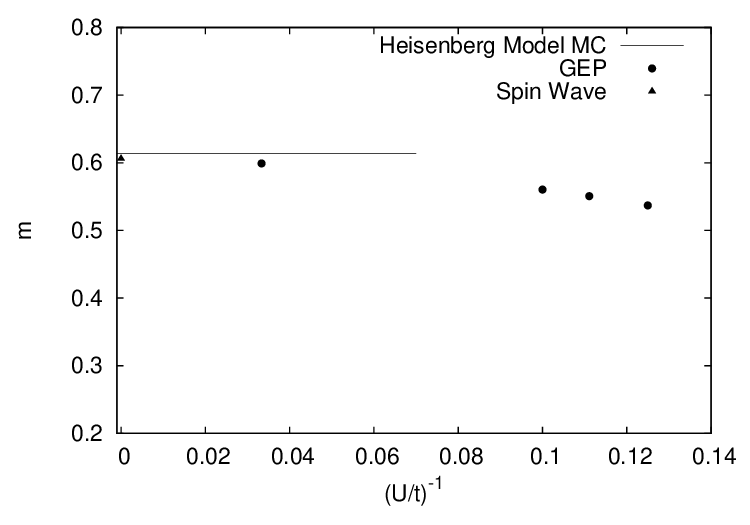}
\caption{magnetization limiting behaviour for the GEP (full circles) as a 
function of the inverse coupling constant $(U/t)^{-1}$. The straight 
horizontal line shows the Heisenberg Model magnetization as obtained 
with a Monte Carlo (MC) by Sandvik \cite{heisenberg_mc} while the triangle 
represents the spin-wave result for the same model \cite{heisenberg_sw}:
they are shown in correspondance to $(U/t)^{-1}$ as in that limit 
Hubbard and Heisenberg model should coincide.
As it is evident from the figure, the GEP magnetization 
trend confirms this expectation, supporting our claim for the GEP 
to be very reliable in the strong coupling limit.
}
\label{fig:3}
\end{center}
\end{figure}

This trend is evident when we compare our results for the 
magnetization with other calculations present in literature. 
The standard for numerical studies on the half-filled Hubbard model
was set by Hirsch's seminal paper of 1985\cite{MC}, which is still
regarded as one of the best reference point on the subject\cite{MCrev}.
Hirsch's work shows that Monte Carlo predictions for the magnetization 
tend to the infinite coupling limit as the ratio $U/t$ increases, 
but the convergence is slower than expected by the RPA\cite{RPA} data
that saturate for $U/t>8$ already, as reported in Fig~\ref{fig:2}. Even if
the Monte Carlo data show large error bars in the strong coupling limit,
the RPA prediction seems to be quite inaccurate in this limit, and out of
the error bars of Monte Carlo data. Definetely the GEP data seem to be more
acurate than RPA in this strong coupling limit, and fall inside the error bars
of Monte Carlo numerical estimates. It is relevant to point out that the GEP
works better just where RPA and Monte Carlo seem to be less accurate.  
As shown in Fig~\ref{fig:2},
for $\coupling$<4, the RPA \cite{RPA} and 
Monte Carlo \cite{MC} predictions converge reasonably well towards 
the mean field approximation, while the GEP gives wrong results 
due to a bad averaging process by a Gaussian functional which
becomes too wide, as discussed in the previous section.
However, as shown in the same figure, 
the approximation improves in the strong coupling region 
($\coupling\geqslant$8). Data show a tendency to saturate to a limit value 
and this trend appears slower than with RPA, in agreement with Monte Carlo 
results. 
In order to get the saturation value we performed a calculation using the very 
strong coupling constant $\coupling$=30, obtaining $m$=0.6 (see also 
figure \ref{fig:subfig:b}). This limit represents a very important 
consistency test for the GEP. It is in fact straightforward to show that 
for large $\coupling$ the Hubbard Hamiltonian \eqref{Hubb_H} reduces to 
the anti-ferromagnetic Heisenberg model with exchange coupling $\frac{2t^2}{U}$.
A comparison between our magnetization and Monte Carlo simulations for 
the Heisenberg model \cite{heisenberg_mc} 
points out that the GEP has the correct limiting behaviour (figure \ref{fig:3}).
This result is also in agreement with spin-wave theory predictions~\cite{heisenberg_sw}.
Moreover, also in the intermediate region our 
results nicely interpolate between the weak and the strong 
coupling limits and are closer to Monte Carlo calculation than 
RPA.
The prediction of a correct strong coupling limit 
tells us that the relevant fluctuations have been 
included by the gaussian approximation. In fact even in the large $U$ limit, 
the quantum fluctuations are important and reduce the local moment by almost
one half. Thus even if the GEP is not exact, the strong coupling limit 
gives evidence that the error is under control, and that the gaussian
fluctuations are enough for describing the strong coupling regime of the model.

In conclusion, we have shown that the GEP can be regarded as 
a useful non-perturbative tool for studying magnetic systems 
with a broken symmetry ground state, and that
it can be regarded as a complementary tool together with
other well established approaches 
like FRG, FLEX or RPA.
In particular, we have been able to evaluate the effective 
potential for the Hubbard model and to write down an analytic 
expression (\eqref{GEP}) where the ``classical'' 
electron and spin energies and the quantum spin fluctuations 
are clearly recognizable.

For the half filled Hubbard model we compared the magnetization
with the result of other approximations showing 
that the GEP provides an improvement over the RPA approach in 
the intermediate and strong coupling limit, while it seems to be 
inaccurate for small values of $\coupling$. However in the
weak coupling limit the failure seems to be a consequence of
the simple second order expansion, and we expect that the exact
GEP should predict the correct magnetization. In this limit an
improvement could be achieved by insertion of higher order
terms in  the expansion \eqref{expansion}.

Further details about the GEP approximation can surely be obtained 
investigating its dependence on the band filling. The method of
section \ref{sectionIII} still holds for any filling of the band,
while the half-filling electron Green's function (\ref{Green_f})
should eventually be replaced by the general one which depends on
the filling of the band. 
That would allow us to construct a phase diagram for the 2D 
Hubbard Model and discuss
the competition between antiferrommagnetic order and 
d-wave superconductivity\cite{RG,Flex}.


\begin{thebibliography} {99}
\bibitem{fradkin2} A more detailed derivation can be found in Fradkin's book 
                    \cite{fradkin}.
\end{thebibliography}
\end{document}